\documentstyle[12pt]{article}
\def\L{${\cal L}$ }
\def\be{\begin{equation}}
\def\ee{\end{equation}}
\def\bea{\begin{eqnarray}}
\def\eea{\end{eqnarray}}
\def\ba{\begin{array}}
\def\ea{\end{array}}
\def\a{\alpha}
\def\b{\beta}

\def\d{\delta}
\def\e{\epsilon}
\def\G{\Gamma}
\def\0{$\Gamma_0$}
\def\l{\lambda}
\def\o{\omega}
\def\Ld{\Lambda}
\def\r{\rho}
\def\t{\theta}

\begin{document}

\baselineskip=0.5cm

\begin{center}

{\Large\bf    Interacting dimers on the honeycomb lattice:
An exact solution of the five-vertex model}
\vskip 1cm

{H. Y. Huang and F. Y. Wu}

{Department of Physics}

{Northeastern University}

{Boston, Massachusetts 02115}

\bigskip
{H. Kunz}

{Institut de Physique Th\'eorique}

{Ecole Polytechnique F\'ed\'erale}

{Lausanne, Switzerland}

\bigskip
{D. Kim}

{Center for Theoretical Physics}

{Seoul National University}

{Seoul, Korea 151-742}

\vskip 1cm
{\bf Abstract}
\end{center}
\medskip
\medskip

The problem of close-packed dimers on the honeycomb lattice
was solved by Kasteleyn in 1963.  Here we
extend the solution
to include interactions
between neighboring dimers in two spatial lattice directions.
The solution is obtained
by using the method of Bethe ansatz and by converting the
dimer problem into a five-vertex problem.
The complete phase diagram is obtained
and it is found that a new frozen
phase, in which the attracting dimers
prevail,  arises  when the interaction is attractive.
For repulsive dimer interactions a new first-order line
separating two  frozen phases
occurs.
The transitions are continuous and
the critical behavior
in the disorder regime is found  to be
the same as in the case of noninteracting dimers
characterized by a specific heat exponent $\a=1/2$.

\newpage

\setcounter{section}{0}
\renewcommand{\theequation}{\arabic{section}.\arabic{equation}}
\setcounter{equation}{0}

\renewcommand{\thetable}{\Roman{table}}
\setcounter{table}{0}
\renewcommand{\arraystretch}{3}

\section{Introduction}

An important  milestone of the modern theory of lattice statistics
is the  exact solution of the dimer problem
obtained by Kasteleyn \cite{k60} and by Fisher \cite{fisher60}.
Kasteleyn and Fisher considered the problem of close-packed dimers
on the simple quartic  lattice
and succeeded in evaluating
its generating function in a closed-form expression.
While the solution shows that close-packed dimers on the
square lattice do not exhibit
a phase transition, Kasteleyn \cite{k64} later
pointed out that dimers
on the honeycomb lattice do possess phase changes,
and that the transitions are
accompanied by frozen ordered states.
The solution, which has since been analyzed by one of us
 \cite{wu68,wu67}, can be used to describe domain walls
in two dimensions \cite{domain,nk94}.

In this paper we
consider once again close-packed dimers on the honeycomb lattice,
but now with the introduction of interactions between
neighboring dimers along two lattice directions.
We show that, with the onset of dimer-dimer interactions,
a new ordered phase emerges if the interaction
is attractive.  For repulsive interactions the phase
diagram is drastically changed and a tricritical point
emerges.
We deduce locations of all phase boundaries and study its
critical behavior.

We analyze  interacting dimers by   first converting
the problem into a five-vertex model.
For noninteracting dimers    this leads to
a free-fermion model \cite{fw70}
which can  be solved by using the method of
Pfaffians \cite{k60,wu68,wu67}.
But when interactions are present the five-vertex model
has general vertex weights and the method of
Pfaffians is no longer applicable.
While its solution
is in principle
obtainable from that of the six-vertex model by
Sutherland, Yang and Yang announced in \cite{syy67},
but details of \cite{syy67}
has not yet been published.
Likewise,  recently published analyses of the general six-vertex model
in the regime $\Delta<1$ by Nolden \cite{nolden} and in the regime
$\Delta \geq 1$ by Bukman and Shore \cite{bukman,shore},
 where $\Delta$ is a parameter
occurring in the six-vertex model, do not readily translate into
the five-vertex problem since the
five-vertex model corresponds to taking the $|\Delta| \to \infty$
limit.  In fact, it is precisely because of this special situation that
 the analysis of the five-vertex model
as a limit of the six-vertex model requires special care.
     To be sure, several authors \cite{nk94,gbl93} have
recently studied the five-vertex model.  But
 the five-vertex model
 considered in \cite{gbl93} is confined to
 a special regime of the parameter
space which does not yield the complete
complexity of the system. The treatment in \cite{nk94},
which was aimed to studying domain walls, is more complete but
analyzes the Bethe
ansatz solution along a line
 somewhat   different from what we shall present,
and is not transparent in extracting relevant information on the dimer
system.
It is therefore useful to have
an alternate and self-contained analysis of the five-vertex
model in the language of dimer statistics.

We take up this subject matter in the present paper.
Our  approach is essentially that of \cite{syy67}, by considering
the solution of the Bethe ansatz equations in the complex plane.
However, we follow the Bethe ansatz solutions closely and
explicitly carry out all relevant contour integrations as dictated
by relevant physical considerations.
This leads  to a complete and clear picture of
the phase diagram and critical behavior of the interacting dimer system.
Particularly, we find the emergence of a new ordered phase for
attractive
dimer-dimer interactions,  and
the existence of
a first-order line terminating at a new kind of
tricritical point, when the interactions are
repulsive.

The organization of our paper is as follows.
The problem of interacting dimers is
defined in section 1 and mapped into a five-vertex model in section 2.
The Bethe ansatz equation
is set up in section
3, and solved in section 4 in the case of noninteracting
dimers.  In section 5 we analyze the general Bethe ansatz equation,
obtaining expressions for the free energy and
its derivative.  This leads to  the determination of the
contour of integrations in section 6 and the complete phase diagram
in section 7.  Finally, the critical behavior is determined
in section 8 by applying perturbation calculations
to the free energy.

\section{The five-vertex model}
\setcounter{equation}{0}

Consider close-packed dimers on an honeycomb lattice \L
which we draw as a ``brick-wall" shown in Fig. 1.
To each dimer
along  the three edges incident at a vertex, one associates
a fugacity, or weight, $u,v$, or $w$.
A  vertical $u$ dimer  and a horizontal $v$
dimer  are said to be
neighbors if they  happen to occupy  two  neighboring
sites in the same row.
  Let two neighboring $u$ and $v$ dimers interact with
  an energy $-\epsilon$ and thus possessing
a Boltzmann factor
\be
\sqrt \l=e^{\epsilon/kT},\label{lambda}
\ee
with $\l>1$ ($\l<1$) denoting attractive (repulsive) interactions.
 Other pairs of dimers such as $u$-$u$, $u$-$w$ etc. are not interacting
 in our model.
 Then, by replacing the two sites inside each dotted box containing
a $w$-edge in Fig. 1
by a single vertex, and regarding a dimer incident to this vertex
as a bond covering the corresponding lattice edge,
the honeycomb lattice \L reduces to  a simple quartic lattice, and
dimer coverings on \L lead to vertex configurations of a five-vertex model
\cite{wu68}.  Configurations of the
 five-vertex model are shown in Fig. 2 in the context of a six-vertex model.
It is  straightforward to verify that we have  the correspondence
\be
\{\o_1,\o_2, \o_3,\o_4,\o_5,\o_6\} = \{0, w, v, u,
  \sqrt{\l uv},  \sqrt{\l uv}\}.\label{weights}
\ee
Here, for definiteness, we assume $\{u,v,w\} >0$.
Note that (\ref{weights})  is the most general five-vertex model,
since one can always take
$  \sqrt{\l uv}=\sqrt{\o_5\o_6}$, if $\o_5\not=\o_6$.

The partition function of a vertex model is  defined as
\be
Z=\sum_{\rm config}\>\prod_{\rm vertices} \o_{k(v)}, \label{pp}
\ee
where the summation is taken over all allowed vertex configurations,
the product is over all vertices of the square lattice and $\o_{k(v)}$
stands for the weight of a vertex $v$.
The case of $\epsilon=0$ or  $\l=1 $
leads to the free-fermion model
satisfying the free-fermion condition $\o_1\o_2+\o_3\o_4 = \o_5\o_6$
\cite{fw70}.

For a simple quartic lattice of size $M\times N$, one defines the per-site
free energy
 \be
f(u,v,w; \l) \equiv \lim_{M,N\to \infty}
{1\over {MN}} \ln Z,\label{fe}
\ee
for the five-vertex model.  It follows that the per-site generating function
for the dimer problem defined in a similar way is $f/2$.

\bigskip
\noindent
{\it Ordered States}:  It is instructive to examine the possible
ordered states of the dimer lattice.  When  $u$, $v$, or $w$
dominates, the ordered states are those shown respectively in Figs.
3a, 3b, and 3c, where the lattice \L is completely covered by
 $u$, $v$, or $w$ dimers.
These are  the ordered states occurring in
the free-fermion case \cite{wu68}.  But when $\l$ dominates
(large positive $\e$), a new ordered state  can materialize as shown in Fig. 3d.
It is this ordering  that adds to new features to the
interacting dimer system.

\section{The Bethe ansatz equation}
\setcounter{equation}{0}

To begin with consider the general six-vertex model on
 a simple quartic lattice of
$M$ rows and $N$ columns with periodic boundary conditions in both directions.
 Applying transfer matrix in the vertical direction and using
the fact that $n$, the number of empty edges (those not covered by bonds)
in a row of vertical edges, is conserved,
 one can evaluate the partition function (\ref{pp}) using the
Bethe ansatz \cite{Lieb67}.  The Bethe ansatz formulation
for the general six-vertex model has been given in \cite{syy67,Lw72} in
  a ferroelectric language from which
the five-vertex limit does not follow straightforwardly. Here, for completeness,
we state  the Bethe ansatz equation for the six-vertex
model  in terms of the vertex weights \cite{gaudin}.

 In the limit of large $M,N$, one finds
\be
Z\sim \max_{\{n\}}\>\> \bigl[\Ld _R(n)+ \Ld_L(n) \bigr]^M, \label{z}
\ee
with
\bea
\Ld_R (n)&=& \o_1^{N-n}\prod_{j=1}^n\biggl(
 { {\o_3\o_4 -\o_5\o_6 -\o_1\o_3z_j }
\over {\o_4-\o_1z_j }} \biggr) \nonumber \\
\Ld_L (n) &=& \o_4^{N-n}\prod_{j=1}^n \biggl(
{ {\o_1\o_2 -\o_5\o_6 -\o_2\o_4z_j^{-1} }
\over {\o_1-\o_4z_j^{-1}}}\biggr), \label{eigenv}
\eea
for $n \leq N$, where the $n$  complex
numbers $z_j,\> j=1,2,...,n$ are the solution of the Bethe ansatz
equation
\be
z_j^N = (-1)^{n+1} \prod_{i=1}^n \biggl(
{{B(z_i, z_j)} \over {B(z_j, z_i)}} \biggr), \hskip 0.5cm j=1,2,...,n,\label{ba}
\ee
with
\be
B(z,z') =\o_2\o_4+\o_1\o_3\>zz' -(\o_1\o_2+\o_3\o_4-\o_5\o_6) z'. \label{bba}
\ee
Note that for fixed $1\leq n \leq N$,
one has generally ${N\choose n}$ different
  $\Ld_R(n)$ and $\Ld_L(n)$.  It is understood that it is the largest ones
for each $n$ that are used in (\ref{z}).
We remark that a useful parameter occurring in the analysis of the
six-vertex model is
$$
\Delta = \frac{\o_1\o_2 + \o_3\o_4 -\o_5\o_6}{2\sqrt{\o_1\o_2\o_3\o_4}}.
$$
It is then clear that  $|\Delta| \to \infty$ in the limit
of $\o_1 \to 0$, making the five-vertex model a very special limit.

Specializing
(\ref{eigenv}) - (\ref{ba}) to the five-vertex model
weights (\ref{weights}), one obtains
\bea
\Ld_R(n) &=& (\b w)^N \d_{n,N} \nonumber \\
  \Ld_L(n) &=& u^N \prod^n_{j=1} (x_1+x_2z_j), \label{eigen5v}
\eea
where the $z_j$'s  are to be determined from the Bethe ansatz equation
\be
z_j^N =(-1)^{n+1} \prod^n_{i=1} \biggl( {{1-\b z_j}\over {1-\b z_i}}
\biggr), \hskip .5cm j=1,2,\cdots,n \label{ba5v}
\ee
with
\be
x_1 = {w\over u}, \hskip 1cm x_2 = {v \over {u}}\l,
\hskip 1cm \b={{v}\over {w}}(1-\l). \label{xxb}
\ee
It is clear that $\Ld_R$ does not contribute  unless $n=N$.
But for $n=N$ the partition function can be
trivially evaluated.
In this case there are no $u$ dimers and  hence each row of \L
is covered completed by $v$ or $w$ dimers and one has
\bea
Z&=&\bigl( w^N + v^N \bigr) ^M, \nonumber \\
f(u,v,w;\l) &=& \max \{ \ln w, \ln v \}. \label{f_at_1}
\eea
Alternatively, one can show from (\ref{eigen5v}) and (\ref{ba5v}) that
$\Ld_R(N) + \Ld_L(N) = w^N + v^N $ from which (\ref{f_at_1})
also follows.
Hence from here on
we consider $\Ld_L$ only.

Combining (\ref{fe}) and (\ref{z}), one has
\be
f(u,v,w;\l) =\max_{\{n\}} f(n), \label{left}
\ee
where
\be
f(n) = \lim_{N\to \infty} {1\over N} \ln \Ld_L(n). \label{feleft}
\ee
The prescription
of the Bethe  ansatz is that, for each fixed $n$, one solves (\ref{ba5v})
for $z_j$.  This   leads to generally many  sets of solutions.
 One next picks the set of solution
which gives the largest $f(n)$ for each $n$.
Then, the free energy (\ref{left}) is given by
the largest $f(n)$ among all $n$.
We shall refer to the set of $z_j$ which gives
rise to the final expression of the partition function
(\ref{left})
the maximal  set, and the prescription of
maximization the maximal principle.

We first point out some immediate consequences of (\ref{ba5v}).
  First,
it is clear that if $z_j$ is solution of (\ref{ba5v}), then
its complex conjugate $z_j^*$ is
also a solution\footnote{This has the consequence that
one must write $z=|z|e^{i\phi}$ with $-\pi \leq \phi \leq\pi$,
implying that   all branch cuts must be taking along the
negative real axis. This observation plays
a major role in ensuing considerations.}
so that the $z_j$'s are distributed symmetrically
with respect to the real axis.
Secondly, for $N=$ even at least, the negations of $\b$
and $z_j$ leave (\ref{ba5v})
unchanged.  Thus, if $z_j$ is a  solution of (\ref{ba5v}),
then $-z_j$ is the solution when  $\b$ is replaced by $-\b$.
Combining these two observations, we find
the solutions for $\b$ and $-\b$ related by a simple reflection
about the imaginary axis.
Finally,
 multiplying the $n$ equations in (\ref{ba5v}), one obtains
  the identity
\be
\biggl[ \prod_{j=1}^n z_j \biggr] ^N=1. \label{pz}
\ee

\section{Noninteracting dimers}
\setcounter{equation}{0}

It is instructive to apply the Bethe ansatz consideration to
the free-fermion case of \cite{wu68}.
 In this free-fermion case we have
  $\{x_1,x_2\}=\{w/u,v/u\}$,
$ \l=1$, $\b=0$, and  (\ref{ba5v}) becomes
\be
z_j ^N = (-1)^{n+1},\hskip 0.5cm j=1,2,...,n .\label{ffroot}
\ee
Thus the  $z_j$'s are on the unit circle $|z|=1$, and can take on
any $n$ of the $N$ roots of (\ref{ffroot}).
 This gives   ${N\choose n}$ eigenvalues $\Ld_L(n)$
as expected.

For fixed $x_1, x_2$, the maximal set of $z_j$
which gives the largest $\Ld_L(n)$  for each $n$
is obtained by choosing the $n$ largest
  $|x_1+x_2z_j|$.
 Now  for any $n$ write
\be
z_j = e^{i\t_j},\hskip 1cm \a=n/N. \label{a}
\ee
In the limit of large $M$, $N$,
the $z_j$'s
 are distributed continuously on the unit circle
with a uniform angular density $N/2\pi$.
For fixed $\a$ and write $z_\a = e^{i\a \pi}$,  the $n$
  largest $|x_1+x_2z_j|$
are those given by the $z_j$'s on the  arc
 of the
circle extending from $z_\a^*$ to $z_\a$ as shown in Fig. 4.
 One then obtains from (\ref{eigen5v}) and (\ref{feleft}) after replacing
$n$ by $\a$ in the argument
\be
f(\a)  =\ln u + {1\over {2\pi i}}
\int_{z_\a^*}^{z_\a} \ln (x_1+x_2z)dz. \label{ffpp}
\ee
  The maximal  free energy is
therefore, after using (\ref{left}),
\bea
f(u,v,w;1) &=& \max_{\{\a\}}f(\a)  \nonumber \\
&=&f(\a_{0}) \nonumber \\
 &=& \ln u +{1\over {2\pi}} \int^{ \a_0\pi}_{- \a_0\pi}
\ln (x_1 + x_2 e^{i\theta} )d\theta,
\label{maxfe}
\eea
where
${\a_0}$ is determined by
 \be
f'(\a_0)=\ln \bigl|x_1+x_2z_{\a_0}\bigr| =0. \label{phaseb}
\ee

If $1, x_1, x_2$, or equivalently $u,v,w$, form a triangle,
we have $0< \a_0< 1$
and $f=f(\a_0)$  analytic in $u,v,w$.
 If $1, x_1, x_2$, or equivalently $u,v,w$, do not form a triangle,
then there are two possibilities.  For $x_1+x_2 <1$ or $w+v<u$,
  we have $|x_1+x_2z_j|<1$ for all $z_j$, and as  a consequence the maximal
set is the empty set or, equivalently,  $\a_0=0$.  This leads to
 \be
f(u,v,w;1)  = \ln u. \hskip 1cm u\geq w+v  \label{fu}
\ee
 For $|x_1-x_2| >1$ or $|w-v|>u$,  we have
$|x_1+x_2z_j|>1$ for all $z_j$, so that we take  the maximal
set $\a_0=1$ and $z_{\a_0} = e^{i\pi}$.
This leads to after carrying out the integration
in (\ref{maxfe})
\bea
f(u,v,w;1) & =&
 \ln u +{\rm max}\ln \bigl\{ x_2, x_1
\bigr\} \nonumber \\
&=& {\rm max} \> \{\ln v,\> \ln w\} . \hskip 1cm u<|w-v| \label{fwv}
\eea
Thus, the phase boundary is
 \be
|x_1\pm x_2| =1, \hskip 0.8cm {\rm or} \hskip 0.8cm
|w\pm v|=u. \label{pb}
\ee
   These results are in agreement  with  \cite{wu68}.
The ordered states (\ref{fu}) and (\ref{fwv}) are the frozen
states shown in Figs. 3a - 3c
 in which the dimer lattice \L is completed covered by $u, v$, or $w$
dimers.
    The transitions are of second order.
 We note, in particular, that the phase boundary (\ref{pb}) is
determined by setting  $f'(\a_0) =0 $ at
$\a_0= 0$, or $\pi$, the two points where
the path of integration in (\ref{ffpp}) either just emerges
or completes a closed contour.
The observation of  this mechanism underlining
the  onset of phase transitions
proves useful in later considerations.

\section{Analysis of the Bethe ansatz equation}
\setcounter{equation}{0}

We now return to the Bethe ansatz equation (\ref{ba5v}).
Define a constant $\overline C(\a,\b)$ by
\be
\biggl[\overline C(\a,\b) \biggr]^N = (-1)^{n+1}\prod_{j=1}^n (1-\b z_j),
\label{c}
\ee
with $\psi_0 \equiv \arg \overline C$ lying in interval $(-\pi/N, \pi/N]$.
Then the Bethe ansatz  equation (\ref{ba5v}) becomes
\be
\biggl[\overline C(\a,\b)\biggr]^Nz_j^N = (1-\b z_j) ^{\a N} , \label{g}
\ee
 and
the $N$th root of (\ref{g}) gives a trajectory $\Gamma$ on which
all solutions  $z_j$ must reside,
\be
C e^{i\psi_0}
 z_j = (1-\b z_j)^\a e^{i\phi_j}, \hskip 0.5cm \phi_j=2\pi j/N, \>\>
j=1,2,...,N, \label{gamma}
\ee
 where $C = C(\a,\b)\equiv |\overline C(\a,\b)|$.
  The trajectory $\Gamma$ is a curve  in the complex $z$ plane
which is symmetric
with respect to the real axis and is given by the equation
\be
C\bigl| z \bigr| = \bigl| 1-\b z\bigr|^\a.
\hskip 1.5cm (\Gamma)  \label{gamma1}
\ee

However, by diagonalizing the transfer matrix explicitly for
$N\leq 18$, Noh and Kim \cite{dong}
have found that for $\beta>1$ the largest eigenvalue assumes the
``bounded magnon" ansatz of Noh and Kim \cite{nk94} in the form of
\bea
z_1&=& {\bar z}_1 \beta^{\alpha N-1} \nonumber \\
z_j &=& {1\over \beta} \biggl(1-{{{\bar z}_j} \over {\beta^{N(1-\a)}}}
 \biggr), \hskip 0.5cm j=2,3,\cdots,n, \label{bmag}
\eea
where $|{\bar z}_j| =1$ for all $j$.  This says that in the
thermodynamic limit of $N\to \infty$ and $\a\not= 0,1$, one root
$z_1$ resides at infinity while all other roots converge to
$1/\beta$.  Using this ansatz, one obtains from (\ref{g})
\be
C=0, \hskip 1cm \a\not= 0,1, \hskip .2cm \beta>1,
\ee
and from (\ref{eigen5v}) and (\ref{feleft})
\be
f(\a) = (1-\a)\ln u + \a \ln v,
\hskip 1cm \a\not= 0,1, \hskip .2cm \beta>1. \label{new}
\ee
While we shall make use of (\ref{new}) to determine the phase boundary,
however, to make our presentation self-contained
 we shall proceed for the time being without using the anzatz
(\ref{bmag})  and the result
(\ref{new}).  It will be seen that one is led to
the same phase boundary for $\beta>1$.

Define
\be
d \equiv C/|\b|, \hskip 1cm d_c\equiv \a^\a (1-\a)^{1-\a}. \label{dc}
\ee
Then, by examining solutions of (\ref{gamma1}) on the real axis, it is
straightforward to verify that
for $\b>0$, $\G$ assumes the topology shown in Figs. 5a - 5c,
respectively for $d>d_c, d=d_c$, and $d<d_c$.
     The topology of $\G$ for $\b<0$ is  deduced from
 those in Fig. 5 by applying  a reflection
about the imaginary axis, and is  shown in Fig. 6.
 Note that the constant $C$ can be determined
 once one point on $\G$ is known.
Particularly, if $\G$ intersects the negative real axis at $x=-R$, where
$R>0$, then we have
\be
C= h(\a, R) \equiv  {{|1+\b R|^\a}\over{| R|}} .  \label{h}
\ee
Note that, since contours can be deformed as long as they
do not cross poles and move along branch cuts, it is the
topology of the contours that is important.

 In the limit of large $N$, the
distribution of zeroes on  $\G$ becomes  continuous.
 Let $\r(z)$ be the density of zeroes
so that $N\int\r(z)dz$  over any interval
of $\G$ gives the number of  $z_j$'s in that interval.
 Then, using (\ref{gamma}) one finds
\be
\r(z) = {1\over {2\pi i}} \biggl( {1\over z} - { \a\over {z-\b^{-1}}}
\biggr).  \label{density}
\ee

Let  $\G_0$ be the segment(s)  of $\G$ symmetric with respect
to the real axis and on which the  maximal
set of $z_j$'s resides.
 For fixed $\a$, we find as in (\ref{ffpp}) - (\ref{phaseb}),
\be
f(\a) = \ln u + \frac{1}{2\pi i}\int_{\G_0}
\biggl(\frac{1}{z} - \frac{\a}{z-\b^{-1}}
\biggr)\ln(x_1+x_2 z) dz ,\label{fea}
\ee
 \be
f(u,v,w;\l) = f(\a_0), \label{fee}
\ee
where $\a_0$ is the value of $\a$ which  maximizes $f(\a)$.

 Note  that the $\a$-dependence of
the free energy (\ref{fea}) for fixed $\a$
now enters through both $\G_0$ and $\r(z)$. However
they obey two
constraints.
First,
the fact that there are $n$  $z_j$'s  implies
 \be
\int_{\Gamma_0}\r(z)dz=
\frac{1}{2\pi i}\int_{\Gamma_0} \biggl({1\over z}
 - \frac{\a}{z-\b^{-1}} \biggr) dz =\a.\label{c2}
\ee
In addition,
 taking the absolute value of (\ref{pz}), one obtains
\be
\prod _{\G_0} z_j = 1, \label{condition}
\ee
which leads to, in the $N\to \infty$ limit,
 \be
\int_{\Gamma_0}\r(z) \ln z \> dz=
\frac{1}{2\pi i}\int_{\Gamma_0} \biggl({1\over z}
 - \frac{\a}{z-\b^{-1}} \biggr) \ln z dz =0.\label{c1}
\ee
For each fixed $\a$,
the two constraints   (\ref{c2}) and (\ref{c1}) together with the
maximal principle of the free energy are sufficient to determine $\G_0$.
Once $\G_0$ is known,
the free energy can be evaluated using (\ref{fea}) and (\ref{fee}).
 In carrying out integrations along $\G_0$, one is aided by the fact that the
path of integration can be deformed as long as it does not cross poles
nor run along branch cuts.
  Care must be taken, however, when \0
intersects  branch cuts.  When this happens,
 integration along \0 can be computed by completing the
contour into a closed loop and
 using the Cauchy residue theorem.

In our discussions below we shall also need to evaluate $f'(\a)$.
Using (\ref{fea}) for  $f(\a)$, generally
the $\a$ dependence comes in through both the path \0 and
the explicit dependence of $\r(z)$ on $\a$.
   Let $\G_0$ consist of  an open path
running continuously  from $z_0^*$ to $z_0$
and, in addition to the open path, possibly
 another  closed contour intersecting the real axis.
Since \0 can be freely deformed except  the terminal
points and the intersection points with the branch cut,
the derivative of ({\ref{fea}) with respect  to $\a$
is derived from
three kinds of contributions.  First, the two terminal
points $z_0$ and $z_0^*$ will move with $\a$.
We write
\be
z_0\equiv R_0e^{i\t}, \hskip 1cm y\equiv \r(z_0) (dz_0/ d\a),
\label{z0}
\ee
and,  due to the fact
that  the expression (\ref{density}) for $\r(z)$ contains
a factor $1/2\pi i$, we have
\be
y^*=- \r(z_0^*) (dz_0^*/ d\a).
\ee
  Then, the contribution to
$f'(\a)$ due to the $\a$-dependence of the terminal
points gives rise to
$y\ln (x_1+x_2z_0) +y^*\ln(x_1+x_2z_0^*)$.  Secondly, if
\0 consists of another closed contour intersecting the branch
cut in (\ref{fea}) at one or two points $z_r<0$, then
since  the integration path can be deformed,
the $\a$-dependence is through the intercepts
$z_r$ only (which  now moves along the branch cut),
the contribution due to $z_r$ can be treated as in the above.
Namely,
we regard two points just  above and below $z_r$ as
two terminal points. This    leads to
a contribution of $2\pi i \sum_r y_r$, where
\be
y_r = \r(z_r) (dz_r/ d\a) = {\rm pure\>\>imaginary} = -y_r^*,
\ee
and this contribution is the same for all branch cuts.
Thirdly, there is a contribution due to
the second term in $\r(z)$ as shown in (\ref{density}).
Combining the three,  one obtains
 \bea
f'(\a)& =&y\ln (x_1+x_2z_0) +y^*\ln(x_1+x_2z_0^*)
\nonumber \\
& &\hskip 0.5cm +2\pi i \sum_r (\pm y_r)   -{1\over {2\pi i}}
 \int_{\G_0} {{\ln(x_1+x_2z)}\over {z-\b^{-1}}} dz, \label{df1}
\eea
where the $\pm$ sign is determined by the orientation of
\0 at the intercepts.

Finally, it is clear from (\ref{fea}) and (\ref{df1}) that,
in carrying out the contour integrations for $f(\a)$
and $f'(\a)$, it is important to determine the location
of the branch point $-x_1/x_2$ relative to the contour $\Gamma_0$.
First,
from the readily verified identity
\be
(1-\b)-{{x_2}\over {x_1 }} = 1-{v\over w},\label{identity}
\ee
one locates  the branch point $-x_1/x_2$ by
\bea
x_1/x_2 &>& (1-\b)^{-1}, \hskip 1cm w>v \nonumber \\
	 &<& (1-\b)^{-1}, \hskip 1cm w<v, \label{branchpoint1}
\eea
We remark that
from the inequality
 \be
x_1/x_2 < |\b|^{-1}, \hskip 1.5cm \b<0. \label{branchpoint2}
\ee
one has (see below) for Fig. 6c the inequality
$R>R_1>|\b|^{-1}>x_1/x_2$, implying that the points
$x=-R, -R_1$ are both on the branch cut.
 We further relate $\sum_r y_r$, $y$, and $y^*$
by taking the derivatives of (\ref{c2})
and (\ref{c1}) with respective to $\a$, obtaining
\bea
y+y^* &=& 1+ {1\over {2\pi i}}
 \int_{\G_0} {{dz}\over {z-\b^{-1}}} \label{df2} \\
y\ln z_0 + y^* \ln z_0^* + 2\pi i \sum_r (\pm y_r) &=&
{1\over {2\pi i}}
 \int_{\G_0} {{\ln z}\over {z-\b^{-1}}} dz. \label{df3}
\eea
It now follows from (\ref{df1}) - (\ref{df3}) that we have
\be
f'(\a) = y\ln \biggl(\frac{x_1}{z_0} + x_2 \biggr)
 + y^*\ln \biggl(\frac{x_1}{z_0^*} + x_2 \biggr)
+{1\over {2\pi i}}
 \int_{\G_0} {{\ln z- \ln(x_1+x_2z)}\over {z-\b^{-1}}} dz,
\label{df}
\ee
provided that $y_r$'s in (\ref{df1}) and (\ref{df3}) are the same,
namely, \0 cuts both branch cuts at the same points.
In section 7 we shall
compute $f'(\a)$ using  (\ref{df})  which
applies to all cases and all $\a$,
including the case that \0 consists of
an open path as well as a closed contour.
The derivative $f'(\a)$ near the phase boundaries
will also be computed in section 8 by analyzing small
perturbations of the free energy.

\bigskip
\section{The contour $\Gamma$ at $\a =0,1-, 1, $ and $1/2$}
\setcounter{equation}{0}

In our discussions we shall need to evaluate  integrals
at $\a_0=0,1-, 1, $ and $1/2$.  It turns out that the contours $\G$
for $\a=1$ and $\a=1-$ should be considered with care.  In this section
we consider the contour $\G$ at these special points.

\medskip
\noindent
(a) $\a=0$. This is the case that \0 begins to emerge with very few
$z_j$'s. The constraint (\ref{condition}) then dictates that
$z_0\sim 1$ and, using (\ref{gamma1}) and (\ref{condition}),
 one finds $\overline C(0,\beta)=1$ and $\G$ the unit circle.

\medskip
\noindent
(b) $\a =1$. This is the case of $n=N$ when one picks all
$N$ $z_j$'s and hence $\Gamma_0=\G$.
One can use   (\ref{g}) to obtain
\be
\biggl[\overline C^N(1,\b)-(-\b)^N \biggr]z_j^N + \cdots -1 =0. \label{a1}
\ee
Using (\ref{a1}), the constraint (\ref{condition})  leads to
the relation
\be
C^N(1,\b) =  \biggl| 1-\b^N\biggr|,\label{cn}
\ee
or, in the thermodynamic limit,
\bea
C(1,\b)&=& 1, \hskip 2cm |\b|<1 \nonumber \\
 &=& |\b|. \hskip 1.8cm  |\b|>1 \label{cc}
\eea
One also finds  from (\ref{dc}) that $d_c=1$ for $\a=1$, and hence
\bea
 d&>& d_c, \hskip 1cm \hskip 1cm  |\b|<1 \label{dd1} \\
&=&d_c, \hskip 1cm \hskip1cm |\b|>1. \label{dd2}
\eea
 The trajectory $\Gamma_0$  is, from  (\ref{gamma1}) with $C=C(1,\b)$,
 \be
\biggl| {1\over z} - \beta \biggr| =1, \hskip 1cm |\beta| <1 \label{circle}
\ee
\be
\biggl| {1\over {\beta z}} - 1 \biggr| = 1, \hskip 1cm |\beta| >1. \label{line}
\ee
The contour (\ref{circle}) for $|\b|<1$
 is a circle of radius  $(1-\b^2)^{-1}$
 centered at  $x=-\b/(1-\b^2)$ on the real axis,
where $x = {\rm Re} (z)$.
Particularly, the circle intersects the real axis at $x=-R_2, R_1$, where
\be
 R_2=(1-\b)^{-1}, \hskip 1cm  R_1=(1+ \b)^{-1}, \label{xx}
\ee
 The contour (\ref{line}) for
  $|\b|>1$ is the vertical  straight line
 $x=(2\b)^{-1}$.

 It is readily verified that by integrating along the contours (\ref{circle})
and (\ref{line}), one has
\bea
{1\over {2\pi i}} \int_\G \r(z) \ln z dz &=& 0, \hskip 1.8cm |\b|<1 \label{correction1}\\
&=& -\ln |\b| \hskip 1cm |\b|>1. \label{correction2}
\eea
Thus, the constraint (\ref{c2}) is not satisfied for
$|\b|>1$,
indicating that the $\a=1$ solution is spurious.\footnote{However,
if one carries out the product
$\prod _{\G_0} z_j $ in the RHS of (\ref{condition}) for finite $N$,
one can show that (\ref{condition}) is satisfied.}
 This leads us to  consider instead the $\a=1-$ solution as a
 limit to $\a=1$.

\medskip
\noindent
(c) $\a=1-$.

 We consider the cases $\b>0$ and $\b<0$ separately.

For $\b>0$,  the contours are those shown in Fig. 5.
In all cases, since the branch cuts are on the
negative axis, the contour  can  be deformed to form
a single closed contour intersecting the real axis at two points
and enclosing the origin.
Let the intersecting point on the negative real axis be  $x=-R$.
One finds from (\ref{c1}) the relation
\be
\ln h(\a, R) =0, \label{b1}
\ee
where $\a=1-$, $h(\a, R)$ is defined in (\ref{h}) and is equal
to $C$.  Thus, one obtains  $C=1$.
In addition, one can solve for $R$ from (\ref{b1})
and  obtains
\bea
R&=&  (1-\b)^{-1}, \hskip 3.1cm  \b<1 \nonumber \\
R &=& |\b|^{\a/(1-\a)}  \to \infty, \hskip 2cm \b>1. \label{r1}
\eea
 Thus, the intercept with the negative axis $R$
for $\b<1$ is the same as that given in
(\ref{xx}).
  Once $C=1$ is known, one can  compute the
  location of other intercept(s)
with the real axis which will be all positive including the
intercept $(1+\b)^{-1}$ given by (\ref{xx}).  
Note that the second line of (\ref{r1}) 
(and what follows)
is what one would obtain without making use of the
bounded magnon ansatz (\ref{bmag}). 
In this case the  precise
location of the other two intercepts (for $\b>1$, see below)
does not concern us since the contour can be deformed freely
in the $x>0$ plane as long as it does not cross the pole at $\b^{-1}$.

 For $\b<0$  the  contours are those shown in Fig. 6.
Consider first
Fig. 6a
  where the contour intersects the
negative real axis at one point at $x=-R$, one finds
again   $C=1$ and the two intercepts (\ref{xx})
with $R=(1-\b)^{-1}$.
In the case of
Fig. 6c where $\G$  intersects
the   real axis at four points as shown with
$-R< -R_1<-R_2<R_3$,
generally the contour cannot be deformed into a single loop due
to  the presence of
the branch
cuts.  But the integration (\ref{c1})
can be carried out as in the above yielding
\be
-\ln h(\a, R_1) +\ln  h(\a, R_2) + \ln h(\a, R) =0, \label{b2}
\ee
where
 \be
h(\a, R) =   h(\a, R_1) =  h(\a, R_2) = h(\a, -R_3)= C. \label{b3}
\ee
The identities   (\ref{b2}) and (\ref{b3}) can hold only  for  $C=1$,
\be
R_1=(|\b|-1)^{-1}, \hskip 1cm R_2= (|\b|+1)^{-1}, \label{b4}
\ee
which is the same as ({\ref{xx}),
and
both $R$ and $R_3$
diverging  as in the second line
of (\ref{r1}).
Note in particular that $R_2$ coincides with $R=(1-\b)^{-1}$ in (\ref{xx})
of Fig. 5a.
 Now we have again from (\ref{dc}) $d_c=1$  for $\a=1-$.
Thus, in contradistinction to
(\ref{cc}), one finds
 $C(1-, \b)=1$ for all $\b$ and therefore
\bea
d= {1\over {|\b|}} &>&d_c, \hskip 1.8cm |\b|<1 \label{ddd1} \\
	 &<& d_c, \hskip 1.8cm |\b|>1. \label{ddd2}
\eea
Thus, the contour $\G$ consists of two loops when $|\b|>1$,
with the outer loop residing in the infinite regime.

We note that
while ({\ref{ddd1}) is the same as ({\ref{dd1}), ({\ref{ddd2})
is different from ({\ref{dd2}).
Finally, one verifies that,
by deforming $\G$ into
a closed contour enclosing the origin but not
 $\b^{-1}$, the constraint (\ref{c2}) is identically satisfied for all  $\b$.

\medskip
\noindent
(d) $\a=1/2$.

 We are primarily interested in the free energy and its derivatives
near a phase transition point which, as in the nonintersecting
case, occurs when the maximal contour either closes or just
begin to emerge. Thus, for $\a=1/2$. we consider the cases
shown in Figs. 5c  and 6c when the contour $\G$ consists
of two loops and \0 is one of the two loops.
But this cannot happen for $\b>0$.
When $\b>0$ the inner loop and portion of the outer loop are
in the $x>0$ half plane.
Then \0 in the maximal solution of the free energy ({\ref{fea}) cannot be
either of the two loops, since some points on the other loop will
have larger values of $\ln (x_1+x_2 z)$.  Hence $\a=1/2$ can
occur only for $\b<0$.  This conclusion is also expected on physical
grounds, that  the $ \a=1/2$ ordered state in Fig. 3d dominates only  when
the interaction between neighboring $u$ and $v$ dimers
is sufficiently attractive, namely, when $\l$ is
sufficiently large or $\b$ sufficiently
negative.

For $\b<0$ the contour $\G$ intersects the
negative real axis (and the branch cut
of $\ln z$) at three points.
 One verifies after some algebra the identity
$$
\int_{\rm inner\>loop} \r(z) dz = \a \ln \bigl|z_0\bigr| \not= \a,
$$
for any $\a$.  Therefore \0 must be the outer loop.

Taking the outer loop as $\Gamma_0$, we have firstly
from the constraint (\ref{c2})
$$
\a = {1\over {2\pi i}} \int_{\G_0}
\biggl({1\over z} - {\a\over {z-\b^{-1}}}
\biggr) dz = 1-\a,
$$
leading to the correct value $\a=1/2$.
The constraint (\ref{c1}) now yields
\bea
0 &=& {1\over {2\pi i}} \int_{\G_0} \biggl({1\over z} - {{1/2}\over {z-\b^{-1}}}
\biggr) \ln zdz  \nonumber \\
&=& \ln R -{1\over 2} \ln \bigl( R+\b^{-1} \bigr),
\eea
or, equivalently,
\be
R^2 - R -\b^{-1}=0.  \label{requation}
\ee
This yields  the solutions
\be
R_\pm = {1\over 2} \biggl( 1\pm \sqrt{1-{4\over {|\b|}} } \biggr). \label{rpm}
\ee
Thus,  we see that, as expected, (\ref{rpm}) has solution only for
sufficiently negative $\b<-4$.

Furthermore, when \0 is a closed contour,
one has from (\ref{h}), (\ref{dc}) and using (\ref{rpm}),
\be
C\biggl( {1\over 2}, \b\biggr) = \sqrt {|\b|}, \hskip 1cm d_c = {1\over 2},
\ee
and
\be
d= {1\over {\sqrt {|\b|}}} < d_c, \hskip 1.4cm \b<-4.
\ee
Thus, $\G$ consists of two loops only when
$\b<-4$.  In this case $\G$ intercepts the real axis at
the four points $-R<-R_1<  -R_2 <R_3$ as shown in Fig. 6c and determined
from $C|x|= \sqrt {|1-\b x|}$.  This leads to
\bea
R &=& R_+, \hskip 3.5cm R_1=R_- \nonumber \\
R_2&=& {1\over 2} \biggl(\sqrt{1+{4\over {|\b|}}} -1\biggr), \hskip 1cm
R_3= {1\over 2} \biggl(\sqrt{1+{4\over {|\b|}}} +1\biggr).  \label{r4}
\eea

\medskip
In summary, we have found that, for $\a=1-$,
we have $C=1$ and $\G$
consists of one loop for $|\b|<1$ which intersects the real
axis at (\ref{xx}), and two loops for
$|\b|>1$ intersecting the real axis at 4 points.
  In the latter case when $\b>1$, the contour can be deformed
into one loop enclosing the origin and
therefore the situation is the same as for $|\b|<1$.
In the case of $\b<-1$ the outer loop resides in the infinite
regime.  For $\b <1$, the intercept of $\Gamma$ on the negative
axis and closest
to the origin is at $x=-(1-\b)^{-1}$.
For $\a=1/2$, one finds $C=\sqrt{|\b|}$ and that, for $\b<-4$,
$\G$ consists of
two loops intersecting the real axis at the four
points given by
(\ref{r4}).

\section{The phase diagram}
\setcounter{equation}{0}

Ideally one would like to proceed at this point to compute the free energy
(\ref{fee}) from which the
complete  thermodynamics of the dimer system can be determined.
However, as this evaluation involves  path integrations which
generally cannot be put into closed forms,
we shall in the next section apply small perturbations to
the free energy near the phase boundaries.
We proceed here to
first determine  the phase boundaries and the phase diagram.

Guided by the analysis of the $\l=1$ solution of section 4,
we expect singularities of the free energy (\ref{fee}) to occur when  either the
maximal path $\G_0$ contains a small emerging segment or when it completes
a closed contour.
As in the case of $\l=1$,  this will happen at
$\a_0=0$ (segment emerging)
and $\a_0=1$  (closed contour)
leading to the ordered states shown in Figs. 3a - 3c.
In addition, we expect  another singularity
of the free energy to occur at $\a_0=1/2$ for $\b<-4$ as
$\G_0$ completes    the outer loop of
two closed contours, leading to the order state shown in Fig. 3d.
 In all these cases
  the free energy $f(\a)$ should be the  maximal solution
at $\a=\a_0=0,1,$ or $1/2$, and
 the phase boundary is given by
$f'(\a_0)=0$ provided that $f(\a)$ is
 concave at $\a_0$.
 Once the contours \0  is known,
both the free energy (\ref{fea}) and its derivative (\ref{df})
can   be evaluated at $\a=0, 1, 1/2$.

Consider first  $\a= 0$.
This is the case that \0  begins to emerge at $z_0=R_0e^{i\t} \sim 1$
and $y+y^*=1$.
It then follows
from (\ref{fea})
and (\ref{df}) that we have, for $\b<1$ at least,
\bea
f(0) &=& \ln u, \nonumber \\
f'(0) &=& \ln \biggl({{w+\l v}\over u}\biggr). \label{f0}
\eea

For $\a=1,$ (or more precisely $\a=1-$)
 and $ 1/2$, both  integrals
(\ref{fea}) and (\ref{df}) can be evaluated
using the contours \0 determined in the preceeding section.
We leave details of the evaluations elsewhere, and collect here
the results.

\medskip
For $\a=1$ one finds, for $\b>1$,
\bea
f(1) &=& \ln v, \nonumber \\
f'(1) &=& \ln (v/ u), \label{bone}
\eea
 and, for $\b<1$,
\bea
f(1) &=& \max \>\{\ln w, \ln v\} \nonumber \\
f'(1) &=&
  \ln \biggl({{w(w-v)}\over {wu-uv(1-\l)}}
\biggr), \hskip 2.4cm w>v \nonumber \\
&=&
 \ln \biggl({{v-w}\over {\l u}} \biggr). \hskip 4.2cm v>w \label{b12}
\eea

\medskip
\noindent
For $\a=1/2$ one finds, for $\b<-4$,
\bea
f\biggl( {1\over 2}\pm\biggr) &=& {1\over 2} \ln (\l uv) \nonumber \\
f'\biggl( {1\over 2}\pm\biggr) &=&
 \ln \biggl[ {{v\l}\over u}\biggl(1-{w\over {v\l R_\mp}}\biggr)^2\biggr], \label{a12}
\eea
where $R_\pm$ is given by (\ref{rpm}).

To ensure that the free energy $f(\a_0)$ is indeed the
maximal solution and that $f'(\a_0)=0$
is a  phase boundary, we need to ascertain that $f(\a_0)$ is a
maximum.  This turns out to be a delicate matter for
$\a_0 = 0,1$, as detailed calculations
show that $f''(\a_0)=0$. However, one can proceed as follows.

First consider the case $\beta > 1$. As discussed
in Sec. 5 and guided by numerical evidence,
the ground state
is given by the bounded magnon solutions (\ref{bmag}) with
the free energy $f(\a)$ given by (\ref{new}).
It follows that the maximal free energy is
\be
f(u,v,w;\l) = {\rm max}\> \{ \ln u, \ln v\}, \hskip 1cm
w<v(1-\l).
\ee
and the phase boundary for $\beta>1$ is $f(0+) = f(1-)$, or 
\be
u=v \hskip 2cm {\rm for}\> \>\> w<v(1-\l). \label{pb1}
\ee
Alternately, one can also arrive at the same phase boundary
(\ref{pb1}) without invoking (\ref{bmag}) and (\ref{new}):
Assuming that $f(\a)$ is either concave or convex in
$0\leq \a \leq 1$, it is easy to see  that  $f'(0)=0$ and
$f'(1)=0$ cannot be the phase boundary for   $\b>1$.
This follows from the
observations that, using (\ref{f0}),
\bea
f'(0)  \geq  0 &\rightarrow& w+\l v \geq u \nonumber \\
&\rightarrow& v>u  \hskip 1.5cm ({\rm using \>}\>\b>1 \>\>
    {\rm or} \>\> w+\l v <v) \nonumber \\
&\rightarrow& f'(1)>0, \label{ff1}
\eea
and using (\ref{bone}),
\bea f'(1) \leq 0 &\rightarrow& v\leq u \nonumber \\
  &\rightarrow&u>w+\l v \hskip 1cm
({\rm again\>\>using} \>\> \b >1) \nonumber \\
&\rightarrow& f'(0) <0. \label{ff2}
\eea
Thus, for $\b>1$, in the former
case, $f'(0) =0$ is not a phase boundary since
$f(0) < f(1)$, and in the latter case $f'(1) =0$
is not a phase boundary since $f(1) < f(0).$
Furthermore, relations (\ref{ff1}) and (\ref{ff2}) 
are consistent with the fact that the maximal free energy
occurs at, respectively, the frozen states $\a_0 =  1$ and $\a_0=0$.
It follows that the accompanying transition is between the two frozen
phases and is thus of first order.

However, the phase boundary are given correctly by
$f'(0)=0$ and $f'(1)=0$ for $\b<1$. This can be seen as follows.

For $f(0)$ to be a maximum we have always $f'(0) \leq 0$, and
for $f(1)$ a maximum we have $f'(1) \geq 0$. We find
that, for $\b <1$ or $(1-\l)v<w$, one has
\bea
f'(0)\leq 0 &\rightarrow&f'(1) <0 \label{ff3}\\
f'(1) \geq 0 &\rightarrow & f'(0) >0. \label{ff4}
\eea
Thus, under the same convexity and concavity assumption,
$f(0)$ and $f(1)$ can indeed be the maximum of the free energy.
Thus, the phase boundaries are $f'(0)=0$ and $f'(1)=0$ and the accompanying
transition is continuous.
Explicitly, the phase boundaries are, at $\a_0=0$,
\be
w+\l v =u \hskip 1cm{\rm for}\>\> (1-\l)v<w
\label{pb2}
\ee
and, at $\a_0=1$,
\bea
\l &=& \biggr({w\over u}-1 \biggl)\biggr({w\over v}-1 \biggl)
\hskip 1cm {\rm for} \>\> w>v \label{pb3} \\
v&=& w+\l u \hskip 3cm {\rm for}\>\> w+\l v > v > w. \label{pb4}
\eea

Finally, for the phase boundary at $\a_0=1/2$, we observe from
(\ref{a12}) that, for $\b<-4$, $f'(\a)$ is
discontinuous at $\a=1/2$ with
\be
f\biggl( {1\over 2}-\biggr) >
f\biggl( {1\over 2}+\biggr) .
\ee
Then $f(\a)$ is a maximum at $\a =1/2$ provided that we have
\be
f\biggl( {1\over 2}-\biggr) \geq 0 \hskip 1cm
{\rm and} \hskip 1cm f\biggl( {1\over 2}+\biggr)
\leq 0.
\ee
The phase boundary is therefore the
borderline cases $f\bigl( {1\over 2}-\bigr) =0$
and $f\bigl( {1\over 2}+\bigr)$
$ =0$ when $f\bigl({1\over 2}\bigr)$ begins to exhibit a maximum.
Now, for $\b<-4$ and writing $v\l -|\b|w=v$, it can be readily verified that
we have
\be
v\l -{w\over {R_\pm }} \geq v >0.
\ee
Thus, the phase boundary at $\a_0=1/2$ is, from (\ref{a12}),
\be
v\l -{w\over {R_\pm }} =\sqrt{\l uv}. \label{pb5}
\ee
Explicitly, (\ref{pb5}) can be written as
\be
uv\l^3 -[w^2+2w(u+v)+u^2+v^2]\l^2
+[2w^2+2w(u+v)+uv]\l -w^2=0,
\label{pb6}
\ee
which reflects its full symmetry with respect to $u$ and $v$.

\medskip
{\it The phase diagram}.
Since the vertex weights (\ref{weights}) are arbitrary to an
overall constant and since there exists an expected
$u, v$ symmetry, it is convenient to consider the phase
 diagram in the parameter space
$u/w,v/w,\l$.
 We have found the existence of five regimes
$W$ ($\a_0=0$), $U$ ($\a_0=1$), $V$ ($\a_0=1$), $\Lambda$
 ($\a_0=1/2$) and $D$ (disordered).
The regimes $U,V,W,\Lambda$ are  phases in which dimers
are frozen in respective configurations of Figs. 3a,b,c,d
with values of $\a_0$ fixed as indicated.
The phase boundaries are given by (\ref{pb1}),
(\ref{pb2}), (\ref{pb3}), (\ref{pb4}), and (\ref{pb1}), leading to
the phase diagrams shown in Fig. 9.

For $\l=1$, the noninteracting case, the phase diagram is
 shown in Fig. 9a, and the  boundaries
separating regimes $W/D$, $U/D$, and $V/D$
are given respectively by
(\ref{pb2}), (\ref{pb4}), (\ref{pb3}).

For $\l>1$
corresponding to
attractive interactions between $u$ and $v$ dimers,
a new ordered phase $\Lambda$  with $\a_0=1/2$ arises.  A typical phase
diagram is shown in Fig. 9b, where, in addition to those
phase boundaries already present in Fig. 9a,  a new
phase boundary (\ref{pb6}) separates regimes $D$ and $\Lambda$.

For $\l<1$ corresponding to repulsive interactions between
$u$, $v$ dimers, a typical  phase diagram is  shown in Fig. 9c.
In addition to the boundaries already present in Fig. 9a,
regimes $U$ and $V$ now share a boundary given by (\ref{pb1}),
namely $u=v>w/(1-\l)$, across which there is a first-order transition.
This leads to the existence of a special transition point
at $u=v=w/(1-\l)$. It is a point where two lines of continuous transition
merge into a first order line and maybe called a kind of tricritical
point. However, it is different from ordinary tricritical points in that
the discontinuity along the first order line does not vanish at that
point.

\setcounter{equation}{0}

\section{The critical behavior and expansions of the free energy
	 near phase boundaries}

In this section we derive  expansions of the free energy
(\ref{fea}) for small deviations near  phase boundaries,
and use the expansions to obtain the critical behavior
in the disorder regime.

The phase boundaries are characterized by $\a \sim 0, 1$ and
$1/2$ and, in all cases  \0 can be deformed into a 
a single trajectory extending from a point
$z_0^*$ to its complex conjugation $z_0$.
Thus, we write
\be
z_0=R_0 e^{i\t}, \hskip 1cm z_0-\b^{-1}=Ae^{i\phi} \label{r0a}
\ee
where $R_0>0, A>0, 0<\{\t,\phi\}<\pi$.
We consider $\t, \phi \sim 0$ or $\pi$, and
there are three cases to consider.

\medskip

\noindent
(a) $\a=0$:
In this case \0 is a small arc of radius $R_0$ extending from
angle $-\t$ to $\t$. 

For  $0<\b R_0<1$, for example, (\ref{c2}) can be written as
\bea
\a &=& {\t \over \pi}-\a {{\phi -\pi} \over \pi}. \label{alpha1}
\eea
Similarly, for the cases $\b R_0>1$ and $\b<0$, we obtain
\be
\a = {\t\over \pi} - \a{\phi\over \pi}.\label{alpha2}
\ee
Furthermore, in all cases  $R_0$, $A$ and $|\b|^{-1}$ form a triangle implying
the relation
\be
R_0 : A : |\b|^{-1} = \sin \phi : \sin \t : |\sin (\phi-\t)|.
\ee
Thus, one obtains
\be
R_0= \frac{\sin \phi}{\b\sin(\phi-\t)}, \hskip 1cm
A= \frac{\sin \t}{\b\sin(\phi-\t)}, \label{ra}
\ee
where $\b$ always has the same sign as $\phi-\t$.
 For given $\a$ and $\b$,
either (\ref{alpha1}) or (\ref{alpha2}) and the expression of $R_0$ in
(\ref{ra})
relates  $R_0$ to  $\t$.
To determine $R_0$ and $\t$ individually, another relation
connecting $R_0$ and $\t$  is needed.  This is
provided by  (\ref{c1}).

For all cases, (\ref{c1}) can be written as
\be
0=\ln R_0 + {1\over 2\pi} \int_0^\t \ln \biggl|
  {{R_0 e^{i\varphi}-\b^{-1}}\over {R_0 e^{i\t}-\b^{-1}}}
  \biggr|^2 d\varphi. \label{dr1}
\ee
Thus, $R_0$ and $\t$ can be generally determined,
although implicitly.

Now we specialize the above consideration to  small $\a$.
When $\a$ is small,  $\t$ is also small. Then,
expanding (\ref{r0a}) and (\ref{dr1}), one obtains
$R_0=1+O(\t^3)$, $A=|1-\b^{-1}|+O(\t^2)$ and
\be
\t \biggl[ 1+ {\a \over {(\b R_0)^{-1}-1} } \biggr]
=\a \pi,\label{small}
\ee
establishing that $\t\sim \a\pi$.

Using (\ref{alpha1}), (\ref{alpha2}) and the small angle expansion
(\ref{small}), one finds after some algebra
that in all cases the
expansion of the free energy (\ref{fea})  at $\a=0$ is
 \bea
f(\a)=\ln u +\a \ln (x_1+x_2) - {{\a^3 \pi^2} \over 6}
      \biggl({{x_1 x_2}\over {(x_1+x_2)^2}}\biggr)+O(\a^4). \label{fff1}
\eea
Note that the corresponding expression Eq. (34) of \cite{nk94}
contains a typographical error.

\medskip
\noindent
(b) $\a=1$:  In this case the contour \0
is almost a closed loop, and can be considered as a closed loop
$\Gamma'$  intercepting the negative real axis at $-R_0$
plus a small arc extending from $z_0^*$ to $z_0$.
Now we have always  $\t \sim \pi$ and, depending on whether
$\phi\sim \pi$ ($\b>0$) or $\phi\sim 0$ ($\b<0$), we have
the two cases to consider. Thus, (\ref{c2})  leads to
\bea
\a &=& 1+{{\t-\pi} \over \pi}-\a {{\phi -\pi} \over \pi}, \hskip 1cm \b>0
\nonumber \\
&=&  1+{{\t-\pi} \over \pi}-\a {{\phi } \over \pi}, \hskip 1.8cm \b<0
\label{aa2}
\eea
We again  find (\ref{ra}) hold for all $\b$, and that, 
(\ref{c1}) leads to
\bea
0 &=&\a \biggl[ \ln R_0 - \ln |\b A| - {1\over 2\pi}
    \int_0^{\pi-\t} \ln \biggl|
    {{R_0 e^{i\varphi}+\b^{-1}}\over {R_0 e^{i(\pi-\t)}+\b^{-1}}}
    \biggr|^2 d\varphi \biggr] . \label{dr2}
\eea

We now specialize to $\a=1-$ and $\t =\pi-$.  Expanding (\ref{dr2}) and
 (\ref{aa2}),  one obtains
 $R_0=(1-\b)^{-1}+O[(\pi-\t)^2]$,
$A=|(1-\b)^{-1}+\b^{-1}| + O[(\pi-\t)^2]$ and
\be
(\pi-\t) \biggl[ 1- {\a \over {(\b R_0)^{-1}+1} } \biggr]
=(1-\a) \pi. \label{aaa3}
\ee
It then follows that, after
using (\ref{identity}), (\ref{aa2}), (\ref{aaa3})
and some lengthy algebra, one arrives at
the expansion valid for all $\b$,
\bea
f(\a)&=&\ln v + (\a-1)\biggl[ \ln \biggl( x_2-x_1 (1-\b)\biggr)
	  + \ln \biggl({x_1 \over x_2}\b +1\biggr)  \biggr] \nonumber \\
     && \hskip2.5cm +{{(\a-1)^3 \pi^2}\over 6}
     \biggl({{x_2 [x_1-\b (x_1 \b + x_2)]} \over
	{(1-\b)^2 [x_2-(1-\b) x_1]^2}}\biggr), \nonumber \\
	&& \hskip4cm {\rm for} \hskip 0.5cm
       R_0\sim (1-\b)^{-1} > x_1/x_2, \nonumber \\
      &=&\ln w + (\a-1)\ln \biggl[x_1-{x_2 \over {1-\b}}
     \biggr]+{{(\a-1)^3 \pi^2}\over 6}
     \biggl[{{x_1  (x_1 \b + x_2)} \over
	{[x_2-(1-\b) x_1]^2}}\biggr], \nonumber \\
     && \hskip4cm {\rm for} \hskip 0.5cm
       R_0\sim (1-\b)^{-1} < x_1/x_2. \label{fff2}
\eea

\noindent
(c) $\a=1/2$: In this case the contour \0
(after some deformation in the case of $\a={1\over 2}+$)
is  a closed loop plus a small arc
running from $z_0^*$ to $z_0$, both intercepting the negative axis at
$- R_0$ for $\a \sim {1\over 2}\pm$.
Since $-R_\pm < \b <0$, we have always
$\t, \phi \sim \pi$. Therefore  (\ref{c2}) yields
\be
\a=1- \a + {{\t-\pi}\over \pi}- \a {{\phi-\pi}\over \pi}.
  \label{c2a}
\ee
One also finds $R_0$, $A$ given by (\ref{ra}).
In addition, (\ref{c1}) now leads to
\be
0=2\ln R_0- \ln A -{1\over {2\pi}} \int_0^{\pi-\t} \ln \biggl|
  {{R_0 e^{i\varphi}+\b^{-1}}\over {R_0 e^{i(\pi-\t)}+\b^{-1}}}
  \biggr|^2 d\varphi .  \label{dr3}
\ee
Thus,   one obtains
$R_0=R_\mp + O[(\pi-\t)^2]$, $A=R_\mp+\b^{-1}+O[(\pi-\t)^2]$, 
\be
(\pi-\t) \biggl[ 1- {\a \over {(\b R_0)^{-1}+1} } \biggr]
= 2 \biggl({1\over 2}-\a\biggr) \pi ,
\ee
from which one deduces after some lengthy algebra the expansion
\bea
f(\a )&=&{1\over 2} \ln (\l uv) + \biggl(\a-{1\over 2}\biggr) \ln
	\biggl( {{(x_1-x_2 R_\mp)^2}\over {x_2 R_\mp^2}}
      \biggr)  \nonumber \\
      &&\hskip1.5cm +\biggl( {{2 (\a-1/2)^3 \pi^2}\over {3 (R_\mp-1/2)^3}}
      \biggr) D(x_1,x_2,\b), \label{fff3}
\eea
where
\be
D(x_1,x_2,\b)=\b^{-1} + {{2 x_1 x_2 R_\mp^4}\over {(x_1-x_2 R_\mp)^2}}
     -{{(x_1+x_2 \b^{-1}) x_2 R_\mp^2}\over {(x_1-x_2 R_\mp)^2}}.
    \nonumber \label{ded}
\ee
Here, the upper (lower) sign pertains to $\a>1/2$ ($\a<1/2$).
This is an extension of the corresponding expressions
Eqs. (58), (68) and (69) of \cite{nk94}.

\medskip
\noindent
{\it The critical behavior}.
We have obtained  expansions of the free energy (\ref{fea})
in the disorder regime
near the phase boundaries $\a_0=0,1,1/2$ to be given by, respectively,
(\ref{fff1}), (\ref{fff2}), and (\ref{fff3}). It is now a simple matter to
verify that, in all cases, the maximal free energy assumes the form
\be
f[\a_0(t)] = f[\a_0(0)] + c(u,v,w,\l)\> t^{3/2}, \label{critical}
\ee
where $t$ is some measure of a small deviation
from the phase boundary in the parameter space,  $c(u,v,w,\l)$
is a function regular in $t$, and $\a_0(t)$
is the value of $\a$ determined from the maximal principle.
Considered as a
vertex model  \cite{wu68},
for example, $t$ can be $|T-T_c|$, where $T_c$ is
the critical temperature.  It then follows from (\ref{critical}) that
the transition is of second order (continuous)
and the  specific exponent is $\a = 1/2$.
This is exactly the same critical behavior
of noninteracting dimers \cite{k64,wu67},
and is also the critical behavior expected from
the  Pokrovsky--Talapov  type transitions \cite{domain}.
In order to check the internal consistency of our
results, however, we shall define $t>0$ by writing
in respective equations for the phase boundary $w\to w(1+t)$,
or $w\to w(1-t)$ to ensure in the disorder regime.
We then expect the resulting expression for $c(u,v,w,\lambda)$ to reflect
a $\{u,v\}$ and $\a_0=\{0,1\}$ symmetry.

To verify (\ref{critical}), we apply the maximal principle
to the free energy (\ref{fff1}), (\ref{fff2}), and (\ref{fff3}).
Consider first  (\ref{fff1}), the expansion of
$f(\a)$ at $\a=0$, for which the phase boundary is
  $x_1+x_2=1$ or $u=w+\l v$.  Near the phase boundary
we write $w=w(1+t)$,
where $t$ is small, and determine $\a_0(t)$ from
\be
f' [\a_0(t)]=\ln \biggl(1+{w\over u}t\biggr) - {{[\a_0(t)]^2 \pi^2}\over 2}
		\biggl({{x_1 x_2}\over {(x_1+x_2)^2}}\biggr)=0 .
\ee
Substituting this $\a_0(t)$ into (\ref{fff1}) and expanding for
small $t$, one obtains
\be
f[\a_0(t)] = \ln u + \frac{2w}{3\pi} \sqrt{ {2\over {uv\l}} } \>t^{3/2}.
	     \label{cff1}
\ee
This leads to (\ref{critical}).

Consider next (\ref{fff2}), the expansion of $f(\a)$ at $\a=1$.
For the first line of (\ref{fff2}), the phase boundary
is (\ref{pb4}) or $v=w+\l u$.
Near the phase boundary we define $t$ by writing
$w\to w(1+t)$ in (\ref{pb4}) and obtain
\bea
f' [\a_0(t)]&=&\ln (1-{w\over {\l u}}t) + {{[\a_0(t)-1]^2 \pi^2}\over 2}
	    \biggl( {{x_2 [x_1-\b (x_1 \b + x_2)]} \over
	     {(1-\b)^2 [x_2-(1-\b) x_1]^2}}\biggr) \nonumber \\
&=&0.
\eea
This leads to
\be
f[\a_0(t)] = \ln v + \frac{2w}{3\pi} \sqrt{ {2\over {uv\l}} } \>t^{3/2}.
	     \label{cff2}
\ee
Note that (\ref{cff1}) and (\ref{cff2}) reflect the expected
$\{u,v\}$ and $\a_0$ symmetry.

For the second line of (\ref{fff2}), the phase boundary
is $x_1-x_2/(1-\b)=1$ or (\ref{pb3}).  Near the phase boundary
we define $t$ by writing $w\to w(1-t)$ in (\ref{pb3})
and obtain
\bea
f' [\a_0(t)]&=&\ln \biggl(1-{{2w-u-v}\over {w-v}}t\biggr)
	     +{{[\a_0(t)-1]^2 \pi^2}\over 2}
	      \biggl({{x_1  (x_1 \b + x_2)} \over {[x_2-(1-\b) x_1]^2}}
	     \biggr) \nonumber \\
	     &=&0.
\eea
This leads to
\be
f[\a_0(t)] = \ln w - t + \frac{2}{3\pi}
	   \sqrt{ {{2(2 w-u-v)^3} \over {\l uv (u+v)}} } \>t^{3/2}.
	   \label{cff3}
\ee
Here the term linear in $t$ comes from the
first term in (\ref{cff3}) and  does not contribute
to the ``specific heat" exponent.

 Finally,
Consider (\ref{fff3}), the expansion of the free energy at $\a=1/2$.
Near the phase boundary
 $v \l - w/R_\pm = \sqrt{\l u v}$ or (\ref{pb6}), we define $t$
by writing $w=w(1-t)$ in (\ref{pb6}). Then $\a_0(t)$ is determined from
\bea
f' [\a_0(t)]&=& \ln \biggl(1-{w\over {R_\mp}} \sqrt{{1\over {\l u v}}}
		  \biggr)
	    +\biggl( {{2 [\a_0(t)-1/2]^2 \pi^2}\over {(R_\mp-1/2)^3}}
	   \biggr) D(x_1,x_2,\b) \nonumber \\
           &=& 0,
\eea
where $D(x_1,x_2,\b)$ is given by (\ref{ded}).
This yields the maximal free energy
\be
f[\a_0(t)] = {1\over 2} \ln (\l uv) + \frac{2w}{3\pi}
	     \biggl( {1\over {\l uv }} \biggr)^{3/4}
	     \sqrt{ {{uv(1-\l)}\over {2(2w+u+v-2\sqrt{uv\l })}}}
	     \>t^{3/2},\label{cff4}
\ee
which reflects the proper $\{u,v\}$ symmetry.  Results
(\ref{cff1}), (\ref{cff2}), (\ref{cff3}) and (\ref{cff4}) now
confirm (\ref{critical}). 
In writing down (\ref{cff1}), (\ref{cff2}), (\ref{cff3}) and (\ref{cff4}),
we have used the respective critical conditions to simplify
the expressions.

\section{Summary}

We have solved the problem of interacting dimers
on the honeycomb lattice by solving the equivalent five-vertex
model using the method of Bethe ansatz.
The free energy is given by (\ref{fea}) and
the maximal free energy by (\ref{fee}) with $\Gamma_0$,
the contour of integrations, subject to constraints (\ref{c2})
and (\ref{c1}).  Phase transitions are then associated with
contours either just emerging or completing a closed loop.
This leads to the determination of the phase boundaries
(\ref{pb1}), (\ref{pb2}), (\ref{pb3}), (\ref{pb4})
and (\ref{pb6}), and the phase diagrams shown in Fig. 9.
We find the occurrence of a new frozen ordered phase for
attractive dimer interactions, and a new first-order line
ending at a tricritical point for repulsive dimer interactions.
We also find, at $\a=1$,
\0 consist of one loop for $|\b|<1$
and two loops for $|\b|>1$ with the outer loop
residing in the infinite regime.
But in the latter case \0 can always  be deformed
into a single loop in computations of the free energy and
its derivative with respective to $\a$,
much simplifying the algebra.
At $\a=1/2$, \0 is found to be
the outer loop of two loops, both of which
in the finite regime, and this occurs only
for $\b<-4$.
We have also evaluated
the free energy in perturbative expansions near the
phase boundary.  This leads to the
determination of the critical behavior in the disorder regime
with the specific heat exponent $\a=1/2$.

\renewcommand{\thesection}{}

\section{Acknowledgements}

One of us (FYW) would like to thank R. J. Baxter for a useful
conversation;
DK would like to thank M. den Nijs for comments and support during
his visit to University of Washington. 
Work by HYH and FYW is supported in part by
NSF grants DMR-9313648 and INT-9207261,
and work by DK is supported in part
by KOSEF through CTP, by MOE of Korea and by NSF grant DMR-9205125.

\newpage

\begin{center}

{\bf Figure captions}

\end{center}

\noindent
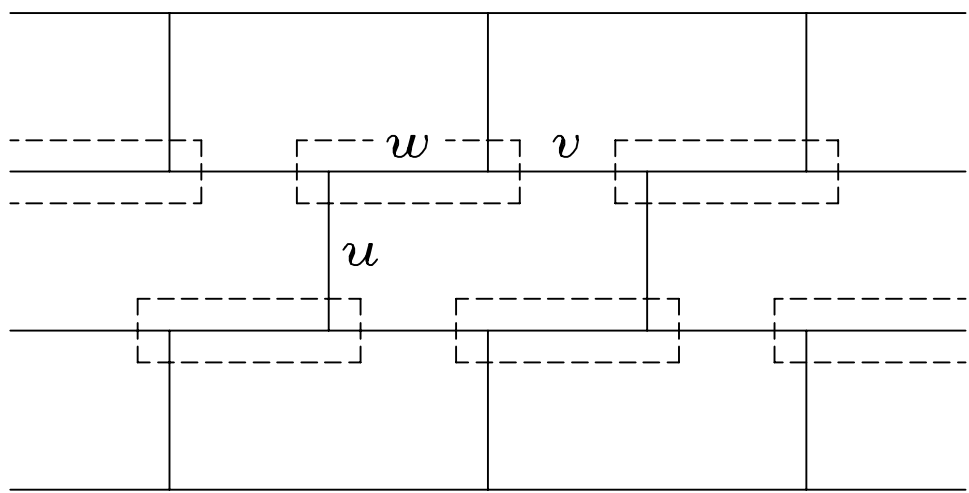
Fig. 1. The honeycomb lattice drawn as a brick-wall lattice showing
relative positionings of the $u$, $v$, and $w$ dimers.
The dotted boxes correspond to vertices of the square lattice.

\medskip
\noindent
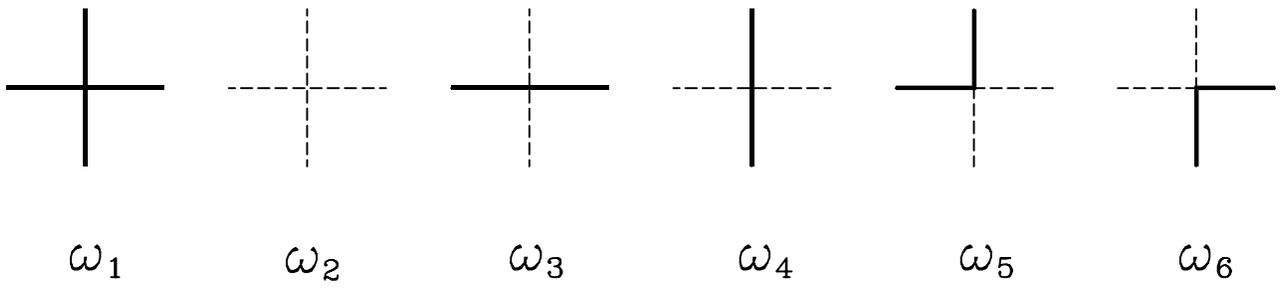
Fig. 2. The six vertex model and the associated weights.

\medskip
\noindent
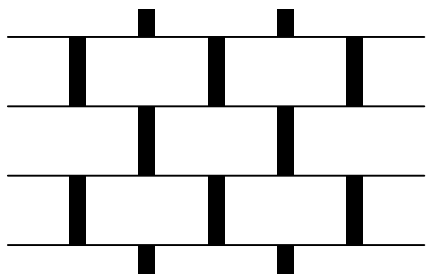
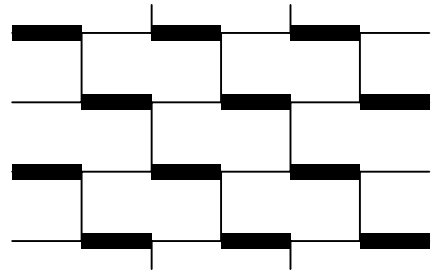
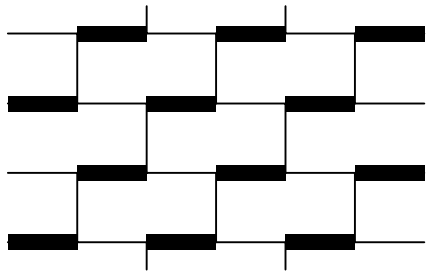
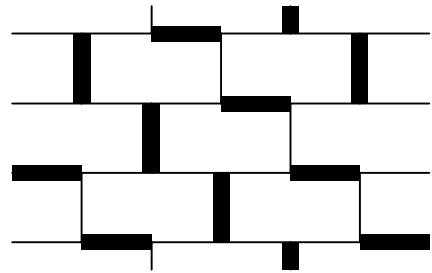
Fig. 3. The four possible ordered states. (a)
The $U$ phase with $n=0$ or $\a=0$ and
$u$ dominating. (b)
The $V$ phase with $n=N$ or $\a=1$ and $v$ dominating.
(c) The $W$ phase with
$n=N$ or $\a = 1$ and $w$ dominating. (d)
The $\Lambda$ phase with $n=N/2$
or $\a = 1/2$ and $\l$ dominating.

\medskip
\noindent
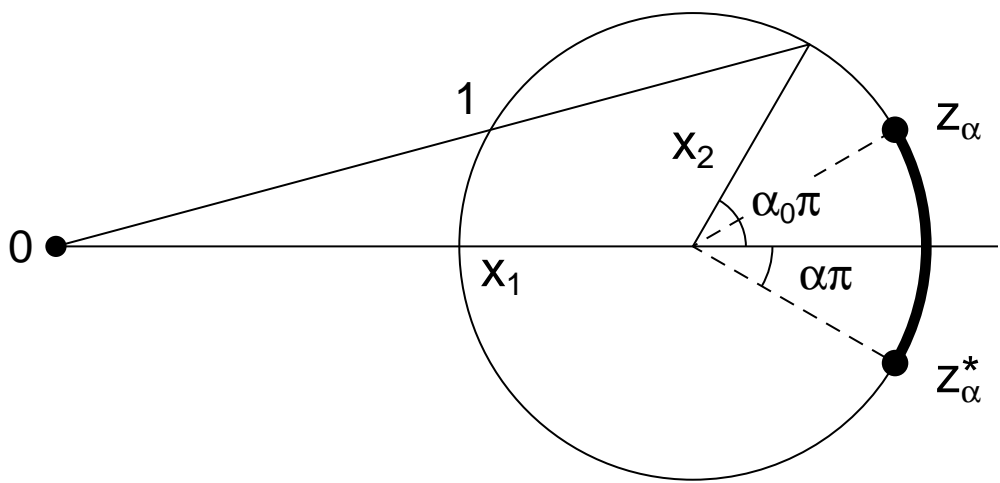
Fig. 4. The maximal contour (heavy curve) for noninteracting dimers.

\medskip
\noindent
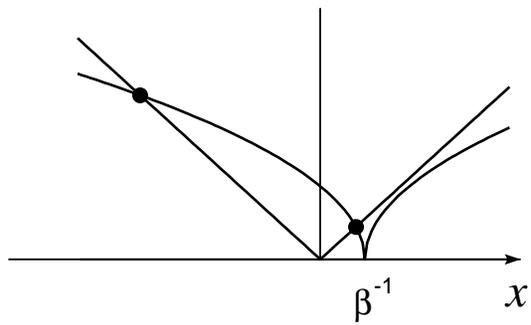
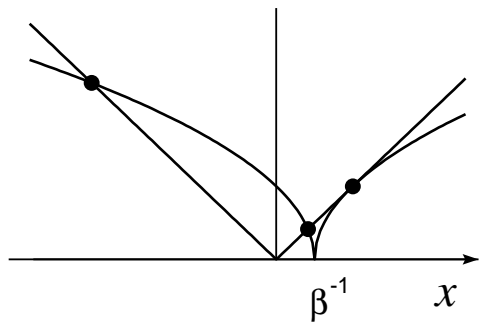
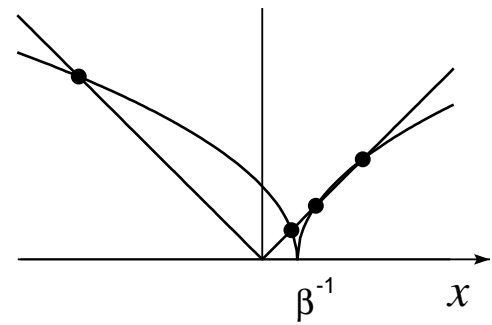
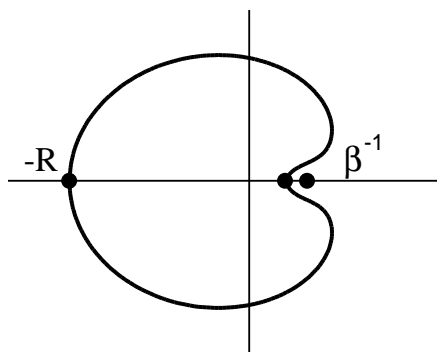
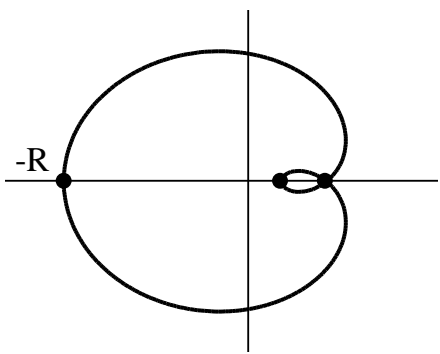
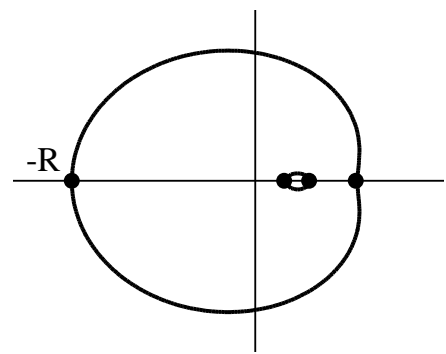
Fig. 5. Possible contours and the corresponding solutions of (\ref{gamma1})
for $\b>0$.
(a) $d>d_c$. (b) $d=d_c$. (c) $d<d_c$.

\medskip
\noindent
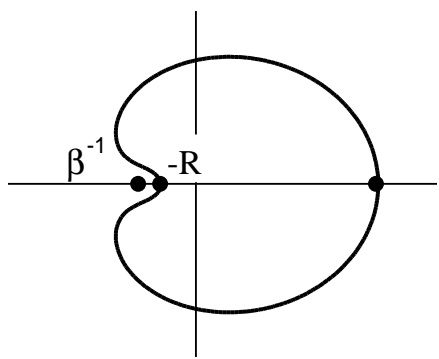
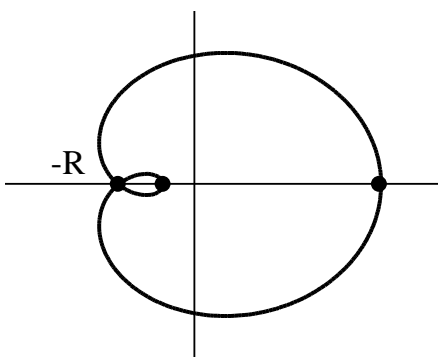
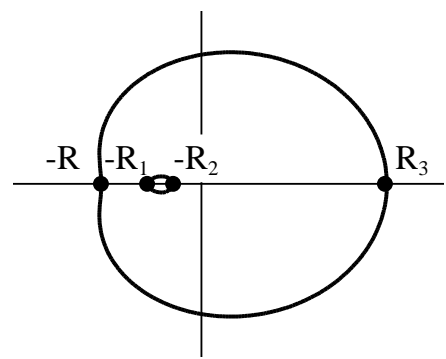
Fig. 6. Possible contours and the corresponding solutions of (\ref{gamma1})
for $\b<0$.
(a) $d>d_c$. (b) $d=d_c$. (c) $d<d_c$.

\medskip
\noindent
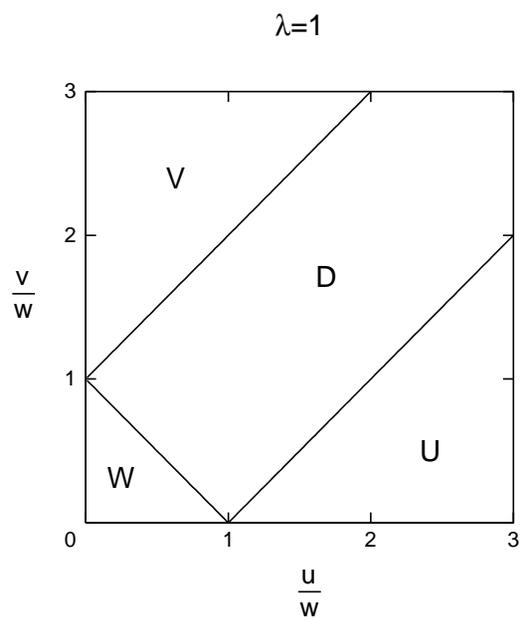
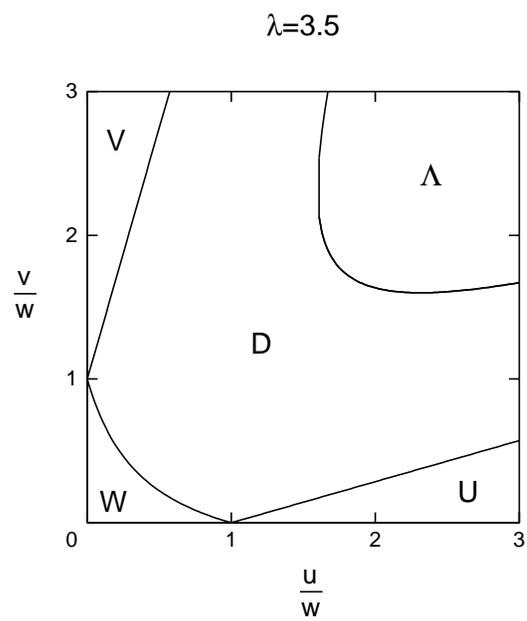
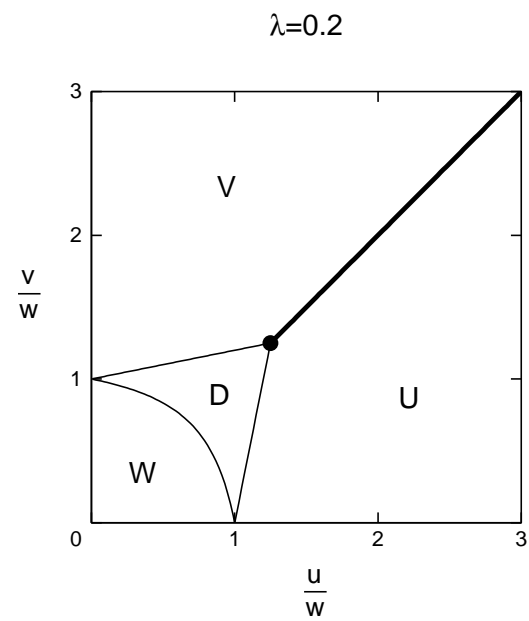
Fig. 7. Phase diagrams for fixed $\l$. (a) $\l=1$, the noninteracting case.
(b) $\l>1$, the case of attractive interactions between
$u$, $v$ dimers.  A new ordered phase $\Lambda$ arises for
$\b<-4$, or $v/w > 4/(\l-1)$. (c) $\l<1$, the case of
repulsive interactions between $u$ and $v$ dimers. The $U$ and $V$
regimes share a first-order boundary denoted by the heavy line,
the circle denotes a tricritical point.

\end{document}